# Path integrals from classical momentum paths

John Hegseth
Department of Physics, University of New Orleans, New Orleans, LA 70148

**Abstract**
The path integral formulation of quantum mechanics constructs the propagator by evaluating the action *S* for all classical paths in coordinate space. A corresponding momentum path integral may also be defined through Fourier transforms in the endpoints. Although these momentum path integrals are especially simple for several special cases, no one has, to my knowledge, ever formally constructed them from all classical paths in momentum space. I show that this is possible because there exists another classical mechanics based on an alternate classical action *R*. Hamilton's Canonical equations result from a variational principle in both *S* and *R*. *S* uses fixed beginning and ending spatial points while *R* uses fixed beginning and ending momentum points. This alternative action's classical mechanics also includes a Hamilton-Jacobi equation. I also present some important points concerning the beginning and ending conditions on the action necessary to apply a Canonical transformation. These properties explain the failure of the Canonical transformation in the phase space path integral. It follows that a path integral may be constructed from classical position paths using *S* in the coordinate representation or from classical momentum paths using *R* in the momentum representation. Several example calculations are presented that illustrate the simplifications and practical advantages made possible by this broader view of the path integral. In particular, the normalized amplitude for a free particle is found without using the Schrödinger equation, the internal spin degree of freedom is simply and naturally derived, and the simple harmonic oscillator is calculated.

**Introduction**
The standard formulation of quantum mechanics assumes probability amplitudes and deduces amazingly accurate quantitative results. A satisfying answer to the qualitative question of "what is a probability amplitude" is still lacking. A very compelling discussion of this amplitude is given in Feynmann and Hibbs [1], where the path integral formulation is presented. This more intuitive formulation looses much of its luster when confronted with basic quantities such as spin. A very basic difference between quantum mechanics and classical mechanics is the failure of Canonical transformations in the phase space path integral. This difference has motivated much interest because of the potential to considerably simplify calculations [2], [3], [4]. Analogous quantum transformations have been recently found [3], [4]. In the following I show how a momentum space path integral may be constructed by considering all possible classical momentum paths. Using a parallel classical mechanics defined using an alternate action *R*, I find all of the necessary features to define a momentum path integral as the sum over all momentum paths of the quantity $e^{i 2\pi \frac{R}{h}}$, where *h* is Planck's constant. In the process of analyzing this classical mechanics, I also show why Canonical transformations do not work in the phase space path integral. This considerably broadened universe of possible path integrals also results in some immediate simplifications. I will demonstrate several specific examples of these practical advantages and simplifications including the calculation of the coordinate path integral normalization constant for a free particle without relying on the Schrödinger equation, a direct and simple calculation of a particle spin propagator, and the simple harmonic oscillator. This new path integral will also show how the momentum representation in quantum mechanics may be viewed as a consequence of this parallel classical mechanics.

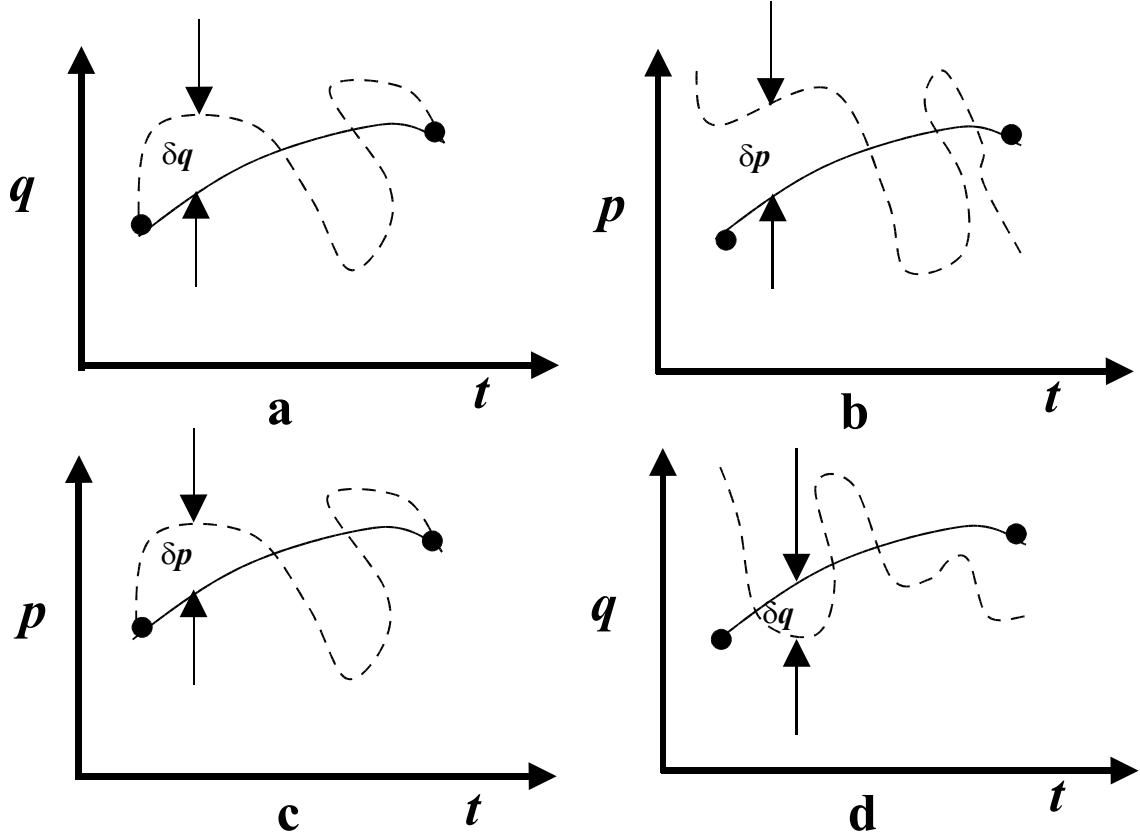

Figure 1. Shown schematically are the paths for *R* and *S* where the solid lines are the critical paths and the dashed lines are arbitrary paths. Figure 1a and 1b shows paths for the action *S[p,q]* where the coordinate path has fixed beginning and ending conditions. Figure 1c and 1d shows the paths for the action *R[p,q]* where the momentum path has fixed beginning and ending conditions.

I now survey some parts of the classical theory that are relevant to this work. Hamilton's' principle tells us that of all the possible paths that a particle may take, the one that extremizes the action $S=\int L[dq/dt,q,t]dt$ is the path that is actually followed [5]. The action *S* has an extrema when $\delta S = 0$, i.e.,

$$\delta S = \int_{t_i}^{t_f} \left( \frac{\partial L}{\partial q} \delta q + \frac{\partial L}{\partial \dot{q}} \delta \dot{q} \right) dt = \int_{t_i}^{t_f} \left\{ \left( \frac{\partial L}{\partial q} + \frac{d}{dt}\left[ \frac{\partial L}{\partial \dot{q}} \right] \right) \delta q \right\} dt = 0.$$

The variation $\delta q(t)$ from the optimal path $q(t)$ is fixed at the initial and final times, i.e., $\delta q(t_i)=\delta q(t_f)=0$ (see Figure 1a). The path that nature selects is the one that makes $\delta S=0$ for arbitrary δq(t) and this implies the Euler-Lagrange equations of motion. The actual path followed requires the selection of two arbitrary constants, e.g., the initial conditions $q(t_i)$ and $\frac{dq}{dt}\big|_{t_i} = \dot{q}(t_i)$. The above is easily generalized to many degrees of freedom. To clarify the notation this discussion is restricted to one.

One important benefit from this classical theory is the flexibility with respect to coordinate transformations that results. Any coordinates that satisfy Hamilton's principle may be used. Even more flexibility is gained if I use the two independent variables *(p, q)*, where the generalized momentum $p = \frac{\partial L}{\partial \dot{q}}$ implicitly defines $\dot{q} = \dot{q}(p,q)$. Introducing the Hamiltonian function *H* the Lagrangian is expressed as [5]

$$L = p\dot{q} - H(p,q). \qquad (1)$$

Asserting the independence of the arbitrary variations δ*p* and δ*q* while only using δq(t_i)=δq(t_f)=0, as shown in Figure 1a and 1b, I can derive Hamilton's equations

$$\delta S = \int_{t_i}^{t_f} \left\{ \frac{\partial L}{\partial q}\delta q + \frac{\partial L}{\partial p}\delta p \right\} dt = \int_{t_i}^{t_f} \left\{ \left( \dot{q} - \frac{\partial H}{\partial p} \right)\delta p - \left( \dot{p} + \frac{\partial H}{\partial q} \right)\delta q \right\} dt = 0 \qquad (2)$$

$$\dot{q} = \frac{\partial H}{\partial p}, \dot{p} = -\frac{\partial H}{\partial q}.$$

In reference [6] it is noted that because of the definition $p = \frac{\partial L}{\partial \dot{q}}$, *p* and *q* are related so that the variations are not independent. The variations δ*q* and $\delta \dot{q}$, being simply related by differentiation when δ*t*=0, are dependent, and the more general relation $p = \frac{\partial L}{\partial \dot{q}}$ means that δ*q* and δ*p* are also dependent. The first Hamilton's equation is then a result of the definition (1). This argument is valid if I take equation (2) to be a consequence of Hamilton's principle expressed in terms of *q* and $\dot{q}$. In the Hamiltonian formulation, however, I generalize *p* to be independent allowing a wider class of possible paths and take equation (2) to be an axiomatic statement. With independent *p, q,* δ*q,* and δ*p,* as shown in Figure 1, both Hamilton's equations are needed to extremize (2); *p* and *q* are connected only by Hamilton's equations [7].

**The momentum action *R***

In the above standard derivation the *p* variable is unconstrained at the beginning and ending points. I may also derive Hamilton's equations using the beginning and ending conditions δp(t_i)=δp(t_f)=0, as shown in Figure 1c and 1d, from the functional *R*

$$R = -\int_{t_i}^{t_f} (q\dot{p} + H)dt = \int_{t_i}^{t_f} Kdt.$$

As before, I allow an independent *p* and *q* and

$$\delta R = -\int_{t_i}^{t_f} \left\{ \dot{p}\delta q + \frac{d(q\delta p)}{dt} - \dot{q}\delta p + \frac{\partial H}{\partial q}\delta q + \frac{\partial H}{\partial p}\delta p \right\} dt$$

$$= -\int_{t_i}^{t_f} \left\{ \left( \frac{\partial H}{\partial p} - \dot{q} \right)\delta p + \left( \dot{p} + \frac{\partial H}{\partial q} \right)\delta q \right\} dt = 0.$$

As can be seen here, δ*R*=0 only if Hamilton's equations are satisfied. There exists another variational principle that uses fixed *p* beginning and ending conditions, as shown in Figure 1c and 1d. In the case of fixed $q$ beginning and ending conditions $\delta q(t_i)=\delta q(t_f)=0$, $R$ is not stationary, i.e., $\delta R= \delta p(t_i)q(t_i) -\delta p(t_f)q(t_f)$. The variations $\delta S$ and $\delta R$ are both zero only when both beginning and ending conditions are satisfied $\delta p(t_i)=\delta p(t_f)=0$ and $\delta q(t_i)=\delta q(t_f)=0$, i.e., fixed initial and final phase space points $\{p(t_i), q(t_i)\}$ and $\{p(t_f), q(t_f)\}$. These points cannot be arbitrarily chosen. If the initial point $\{p(t_i), q(t_i)\}$ is selected, then the two first order Hamilton's equations uniquely determine a path. This later condition can only be satisfied for special phase space points; the points on the actual path at $t_f$. In fact, it is well known that selecting $q(t_i)$ and $q(t_f)$ (corresponding to the beginning and ending conditions $\delta q(t_i)=\delta q(t_f)=0$) is often not enough to specify an actual path between two points, as happens when a path passes through a conjugate point (a point where two or more extremals intersect). The initial momentum is then needed to select one particular extremal from a family of extremals.

The functional $R$ is related to $S$ through partial integration $S(p,q) = R + [pq]_{t_f} - [pq]_{t_i}$ or equivalently $L = K + \dfrac{d(pq)}{dt}$ (3).

Equation (3) may be formally regarded as a Legendre transform from the variables $q$ and $\dot{q}$ to the variables $p$ and $\dot{p}$, with $K$ as the new potential. I relate $K$ to $L(q,\dot{q},t)$ by using the definition $p = \frac{\partial L}{\partial \dot{q}}$ and Lagrange's equation $\dot{p} = \frac{\partial L}{\partial q}$, i.e., $dL = \dfrac{\partial L}{\partial t}dt + \dfrac{\partial L}{\partial q}dq + \dfrac{\partial L}{\partial \dot{q}}d\dot{q} = \dfrac{\partial L}{\partial t}dt + \dot{p}dq + pd\dot{q}$. Taking he total differential of (3) gives: $dL = \dfrac{\partial L}{\partial t}dt + \dot{p}dq + pd\dot{q} = dK + \dot{p}dq + qd\dot{p} + \dot{q}dp + pd\dot{q}$

or $dK = \dfrac{\partial L}{\partial t}dt - qd\dot{p} - \dot{q}dp.$ (4)

Equation (4) shows that $K$ is a function of $p$, $\dot{p}$, and $t$. Comparing equation (4) to the total differential of $K = K(p,\dot{p},t)$ shows that $\dfrac{\partial K}{\partial t} = \dfrac{\partial L}{\partial t}$, $\dot{q} = -\dfrac{\partial K}{\partial p}$, and $q = -\dfrac{\partial K}{\partial \dot{p}}$. The second equation corresponds to the Euler-Lagrange equation for $R$ while the third equation defines the position variable in analogy to $p = \frac{\partial L}{\partial \dot{q}}$. Hamilton's equations may easily be derived by comparing the differential $dH$ to $dK$ (and using $H = -q\dot{p} - K$). A few simple examples, however (e.g., a simple harmonic oscillator), shows that these equations of motion simply repeat the equations that define the variables and nothing new is found (as should be expected since one of the equations that defines a new variable is the equation of motion). These equations also give absurd results in 1-D constant force motion. Nevertheless, this does show that $K$ (like $L$) may also be considered a functional of only one independent function. $K$ is only useful, however, when it is considered a function of both $p$ and $q$.

Because $S$ and $R$ may both be considered functionals of two independent functions, I may gain greater flexibility when transforming variables. These transformations, called Canonical transformations, can be written as $P=P(p, q)$ $Q=(p, q)$. Such a transformation is allowed, however, only if $\delta S=0$ (or $\delta R=0$) is still satisfied in terms of $(P, Q)$. The resulting variation

$$\delta S = \delta \left\{ \int_{t_i}^{t_f} \left( P\dot{Q} - H^*(P,Q) \right) dt \right\} = 0$$

gives Hamilton's equations in *P, Q,* and *H\** if I require $\delta Q=0$ at the beginning and ending points (because the integration by parts must also be done in this case). Note, however, that $\delta Q = \frac{\partial Q}{\partial p}\delta p + \frac{\partial Q}{\partial q}\delta q$ requiring that either $\delta p=0$ or $\partial Q/\partial p=0$ at the beginning and ending points. Because the beginning and ending points occur at arbitrary time, the latter condition is only possible if $Q=Q(q)$ and not in this wide class of transformations. The Canonical transformations require both exact beginning and ending conditions for *p* and *q*. If I further require $\delta p(t_i)=\delta p(t_f)=0$, then this wider class of coordinate transformations, the Canonical transformations, are possible.

**Path integrals and the momentum representation**
In non-relativistic quantum theory all paths are considered possible with equal amplitude and a phase that is proportional to the action for the given path [1]. The probability amplitude for a classical path φ is calculated directly from a classical Lagrangian using the formula $\varphi=exp[i2\pi S/h]$, where *h* is a characteristic quantum of action (Planck's constant). The usual initial and final conditions are defined with respect to space coordinates $q(t)$, i.e., $q(t_i)$ and $q(t_f)$. Summing over all possible paths, or integrating with an appropriately defined measure, yields the path integral that is identified as the propagator between $q(t_i)$ and $q(t_f)$. In other words, I can use *S* expressed in space coordinates to define propagators in coordinate space as is commonly done. Because this propagator propagates a wave function in time between the beginning and ending points of the path integral, it implicitly gives the Schrödinger wave equation. The endpoints of the path integral correspond to the space variables in quantum mechanics. This formulation, however, has many difficulties such as defining the measure of the integral and describing the spin variables of a quantum system. Propagators may also be simply defined in terms of the quantum formalism, without some of these problems, as the matrix elements between different states at different times [8]. In terms of position states it is $G(q_f, t_f; q_i, t_i)=<q_f,t_f|q_i,t_i>$, i.e., the probability amplitude to start at position and time $(q_i, t_i)$ and go to $(q_f, t_f)$. Using the Trotter product formula $e^A= lim_{N\to\infty} (e^{A/N})^N$, where *A* is an operator and *N* is an integer, I can use the time evolution operator to express $G(q_f,t;q_i)=<q_f,t_f|q_i,t_i>=<q_f|e^{-i2\pi Ht/h}|q_i>= lim_{N\to\infty}<q_f|e^{\varepsilon NT}e^{\varepsilon NV})|q_i>=lim_{N\to\infty}<q_f|e^{\varepsilon T}e^{\varepsilon V}e^{\varepsilon T}e^{\varepsilon V}$ ... $e^{\varepsilon T}e^{\varepsilon V}e^{\varepsilon T}e^{\varepsilon V}|q_i>$, where I have set $t = t_f - t_i$, $\varepsilon = \frac{-i2\pi t}{Nh}$, and $H=T+V=p^2/2m+V(q)$. This allows the time interval *t* to be broken into *N* intervals or time slices of duration *t/N* and these time slices go to zero in the limit. I can put this into the path integral form by alternately placing the space and momentum representations of the identity operator, $1=\int dp_j|p_j><p_j|$ and $1=\int dq_j|q_j><q_j|$, between each pair of operators where *j* refers to the *j*[th] time slice. Using the integral $<q|p>=(h)^{-1/2}exp(i2\pi pq/h)$ results in the propagator:

$$G(q_f,t;q_i) = \lim_{N\to\infty} \int dp_1 dq_1 dp_2 dq_2 ... dp_N (h)^{-N} \exp\left\{\varepsilon \sum_{j=1}^{N}\left(\frac{p_j^2}{2m}+V(q_{j-1})\right) + \frac{i2\pi}{\hbar}\sum_{j=1}^{N} p_j(q_j - q_{j-1})\right\}.$$

Because the argument in the exponential is the action for infinitesimal ε, i.e.,

$$\varepsilon \sum_{j=1}^{N} \left( \frac{p_j^2}{2m} + V(q_{j-1}) \right) + \frac{i2\pi}{h} \sum_{j=1}^{N} p_j (q_j - q_{j-1}) = \frac{i2\pi}{h} \int_0^t \left( p \frac{dq}{dt} - H \right) d\tau = \frac{i2\pi}{h} S, \; G(q_f, t_f; q_i, t_i)$$ can be written in the form

$$G(q_f, t; q_i) = \lim_{N \to \infty} \int dp_1 dq_1 dp_2 dq_2 \ldots dp_N \, (h)^{-N} \exp\left\{ \frac{i2\pi}{h} \int_0^t \left( p \frac{dq}{dt} - H \right) d\tau \right\}.$$

This integral can be interpreted as a phase space path integral $\int DqDp e^{i2\pi S/h}$ over all possible paths in phase space between $q_i$ at time $t_i=0$ and $q_f$ at time $t$, where $Dq$ and $Dp$ are appropriately defined measures for the integral. Because of the form of the action given above, the integrals in $p$ can be explicitly performed resulting in the usual form of the position space path integral including the correct normalization amplitude.

Unfortunately the classical Canonical transformations of the above action only produce inconsistencies and not the great simplifications they produce in classical mechanics. I have shown above that a Canonical transformation requires that both $\delta q(t_i) = \delta q(t_f) = 0$ and $\delta p(t_i) = \delta p(t_f) = 0$. In quantum mechanics, however, it is impossible to know $p$ and $q$ exactly at the same time, i.e., this violates the Heisenberg uncertainty principle and explains why they do not work for phase space path integrals. If I define a specific $q(t_i)$ and $q(t_f)$ then I must include all possible $p(t_i)$ and $p(t_f)$ in the integrals at $t_i$ and $t_f$.

I now show that the momentum space path integral follows naturally from the action $R$ (c.f. reference [8]). I do the same analysis in initial and final momentum states, i.e., $G(p_f, t_f; p_i, t_i) = \langle p_f, t_f | p_i, t_i \rangle$ to get the probability amplitude to start at momentum and time $(p_i, t_i)$ and end at $(p_f, t_f)$. In almost the same steps as above, except for the ordering of the $\langle q | p \rangle \langle q | p \rangle$ factors, I get $G(p_f, t; p_i) = \lim_{N \to \infty} \langle p_f | e^{\varepsilon T} e^{\varepsilon V} e^{\varepsilon T} e^{\varepsilon V} \ldots e^{\varepsilon T} e^{\varepsilon V} e^{\varepsilon T} e^{\varepsilon V} | p_i \rangle$ and finally

$$G(p_f, t; p_i) = \lim_{N \to \infty} \int dp_1 dq_1 dp_2 dq_2 \ldots dq_N \, (h)^{-N} \exp\left\{ \varepsilon \sum_{j=1}^{N} \left( \frac{p_{j+1}^2}{2m} + V(q_j) \right) - \frac{i2\pi}{h} \sum_{j=1}^{N} q_j (p_{j+1} - p_j) \right\}.$$

I note that the argument in the exponential is the action $R$ for infinitesimal $\varepsilon$, i.e.,

$$\varepsilon \sum_{j=1}^{N} \left( \frac{p_{j+1}^2}{2m} + V(q_j) \right) - \frac{i2\pi}{h} \sum_{j=1}^{N} q_j (p_{j+1} - p_j) = -\frac{i}{h} \int_0^t \left( q \frac{dp}{dt} + H \right) d\tau = \frac{i2\pi}{h} R.$$

This integral can be interpreted as a phase space path integral $\int DqDp e^{i2\pi R/h}$ over all possible paths in phase space between $p_f$ at time $t$ and $p_i$ at time $t_i=0$, where $Dq$ and $Dp$ are appropriately defined measures for the integral. I can then get a pure momentum space path integral by performing the integrals over $q$ in a similar manner to the position space path integral. In this case a specific potential function is needed to proceed.

**Momentum path integrals from classical paths**

The above discussion suggests that a momentum space path integral can be constructed from classical momentum paths using the classical action $R$ by considering all the possible momentum paths between $(p_i, t_i)$ and $(p_f, t_f)$ as is done in coordinate space in reference [1]. By assuming two complimentary path integrals, a coordinate path integral and a momentum path integral, position and momentum variables are placed on an equal footing. This view also directly traces canonically conjugate variables in quantum mechanics to classical mechanics. I now demonstrate with several elementary examples how this expanded universe of possible path integrals resolves some of the difficulties in the path integral formulation of quantum mechanics.

This view also produces a prescription for treating more complex Hamiltonians where the momentum variables do not separate. This is accomplished by noting that the classical paths in momentum and coordinate space exist in one-to-one correspondence connected only by Hamilton's equations. Each possible $q(t)$ has a corresponding $p(t)$ that may be found by solving $\dot{q} = H_p$ (note that $p$ and $q$ do not refer to critical paths but to arbitrary paths, I use the notation $\Pi$ and $\Theta$ in Appendix B for arbitrary paths). In quantum mechanics $p(t)$ and $q(t)$ cannot be known simultaneously, i.e., a complete knowledge of the position at a given time $t$ implies complete ignorance of the momentum at $t$. At an initial time and position all momenta are possible and all positions are possible at the end of the next time increment. Summing over all possible paths is then equivalent to integrating over all possible positions after each time increment except at the given final position and time. In Appendix B, I show that position paths may specify $S$. Specifically, by using one of Hamilton's equations $\dot{q} = H_p$ to eliminate $p(t)$, I may write $S(p(q), q, t)$ where $p(q)$ solves $\dot{q} = H_p$. In this way the action is treated as only depending on one of the Canonical variables. An elementary example comes from the separable class of Hamiltonians of the form $H = p^2/2m + V(q)$. Hamilton's equations yield $\dot{q} = H_p = \frac{p}{m}$, and in this case replacing $p_j$ by $m\Delta q_j/\Delta t$ become increasingly accurate as $\Delta t \to 0$. This is essentially the standard procedure used in calculating many propagators using coordinate paths. This same reasoning also applies to the momentum paths where each $q(t)$ is replace by $q(p)$ that is a solution to $\dot{p} = -H_q$. The variable $q$ is then replaced in $R$ and the action $R(p,q(p),t)$ is calculated for all possible $p$ paths between $(p_i, t_i)$ and $(p_f, t_f)$. I now illustrate this procedure with several elementary examples.

**The free particle**

The first and simplest example to use the above prescription for the momentum path integral is the free particle. The free particle's Hamiltonian is $H=p^2/2m$, so that $\dot{p} = 0$ or $p=constant$ at every time interval. Summing over all possible momentum paths or integrating over all possible momenta at each time except at the beginning and ending points where $p$ is given constructs the propagator. As shown in Figure 2, I take the initial momentum to be $p_i$ at time $t_i$. At time $t_1$ there are in principle many possible values of momentum. The solution to $\dot{p} = 0$, however, only allows one possibility in this case, and the $p_1$ integral at time $t_1$ is $\int dp_1 \delta(p_i-p_1)\theta(t_i-t_1)$. The $\delta$ function restricts the possible values of $p_1$ to the possible values in the integration, and the $\theta$ function maintains causality. Extending this to all $N$ intervals I get

$$G(p_f, t; p_i) = \lim_{N \to \infty} A \int dp_1 \delta(p_i - p_1)\theta(p_1 - p_i) dp_2 \delta(p_1 - p_2)\theta(p_2 - p_1)...$$

$$... dp_{n-1} \delta(p_{n-2} - p_{n-1})\theta(p_{n-1} - p_{n-2}) \delta(p_{n-1} - p_f)\theta(p_f - p_{n-1}) \exp\left\{\varepsilon \sum_{j=1}^{N-1} \frac{p_j^2}{2m}\right\}.$$

In the above integral each interval must have one $\delta$ and one $\theta$ factor to limit the possible *p* paths and maintain causality. Note that there is one more δ than there are *p*. Integrating this becomes

$$G(p_f, t_f; p_i, t_i) = \lim_{N \to \infty} A\delta(p_i - p_f)\theta(t_f - t_i)\exp\left\{-\varepsilon N \frac{p_i^2}{2m}\right\}.$$

This is the same result given in Feynmann and Hibbs [1] but here it is deduced by summing over all possible momentum paths. The normalization factor has been purposely left ambiguous to demonstrate that in this extended formulation the normalization factor can be found without resorting to the Schrödinger equation (or propagation over a differential in time). As shown in reference [1] and [8] the momentum and position propagators are related by Fourier transformation over the endpoints (this is essentially because of the endpoint terms that relate *R* to *S*). By Fourier transforming the above, I can find the position propagator amplitude to within a constant of *1/A* that is identified, through many independent measurements, as Planck's constant *1/A = h*.

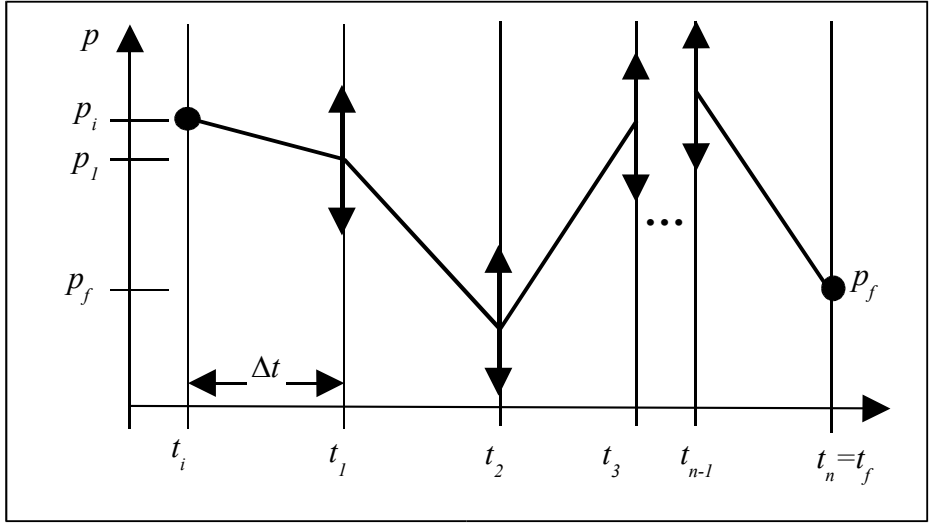

Figure 2. The momentum path integral is constructed by considering all possible momentum paths. This is equivalent to integrating over all possible momenta at each time except at the beginning and ending times.

**The harmonic oscillator**

The Hamiltonian for a harmonic oscillator $H = p^2/2m + \frac{1}{2}m\omega_0^2 q^2$ is symmetric in *p* and *q*, and the allowed Hamilton's equation yields $\dot{p} = -H_q = m\omega_0^2 q$. Substituting for *q* I get $H = \frac{p^2}{2m} + \frac{\dot{p}^2}{2m\omega_0^2}$. In this case, replacing $\dot{p}$ by $\Delta p_j/\Delta t$ becomes increasingly accurate as $\Delta t \to 0$. One can see that this problem is equivalent to the problem of the quadratic Hamiltonian path integral in coordinate space. Many methods have been developed to calculate a path integral of this form and the result is similar. The interested reader can consult references [1], [2], [8], and many others. Although no great advantage is gained in doing a momentum path integral in this case, it is instructive to see how the above prescription is implemented. In this case there is an external force, so that the total uncertainly in position at each time implies that all momenta are possible after the initial time where the initial momentum is given.

**Non-relativistic spin**

One of the advantages of this extended formulation of the path integral is that simple spin variable problems become more tractable. Because the coordinate or angle variables are not observable, even in principle, I immediately go to the momentum propagator because the generalized momentum *is* observable through its association with a particle's magnetic moment. In one dimension, an operational and conceptual simplification occurs as shown in the following. The Hamiltonian for a single spinning particle is $\frac{1}{2}I\dot{\phi}^2$ or $H = \frac{1}{2I}l^2$. As in the case of the free particle $\dot{l} = 0$ or the angular momentum $l$ is constant during each time interval. This case is quite different from linear momentum; linear momentum could be observed by measuring a change in position $\Delta q$ during a change in time. Because $\Delta q$ can be observed to be + or -, the direction of $p$ can be determined. No analogous technique is available here. Because the conjugate variable for angular momentum *is not* observable, even in principle, it is not possible to find the direction of $l$. The value of $l$ is constant but the direction or sign is unknown and has two possible directions along the spin axis $\pm l$. The propagator can then be written as $G(l_f, t; l_i) = \sum_{all\,paths} \exp\left\{-\int_{t_i}^{t_f} \frac{i2\pi}{2Ih} l^2 dt\right\}$. In constructing a possible path, $l$ is taken to be constant and equal to the initial $l$ over a time interval as in the linear momentum case. Figure 3 shows one example of a momentum space path where $l$ can be either $+l$ or $-l$ during each time interval. If there are $N$ such intervals then there are $2^N$ paths so that $G(l_f, t; l_i) = \lim_{N\to\infty} C 2^N \exp\left\{\frac{-i2\pi}{2Ih} l^2 t\right\}$. In order to have a finite propagator, the normalization constant is $C=1/2^N$.

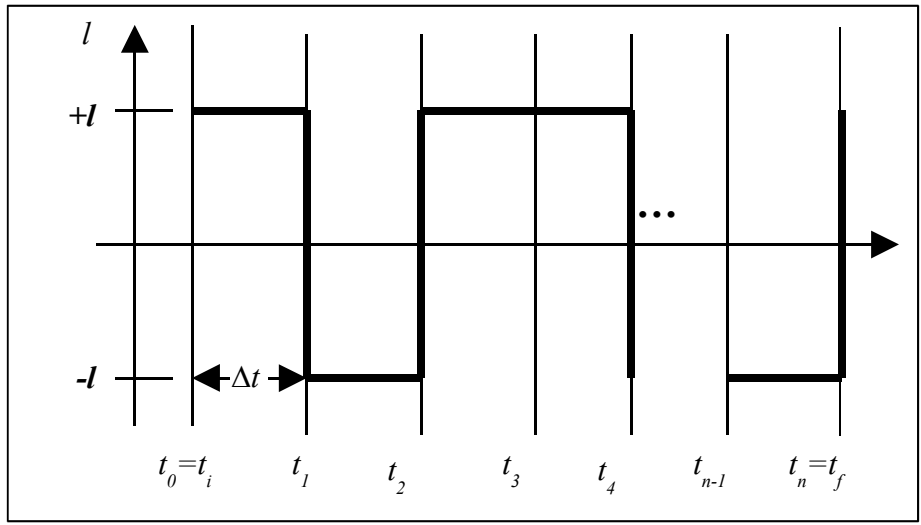

Figure 3. The momentum path integral for the spin of a free particle is constructed by considering all possible momentum paths. The momentum $|l|$ is constant for a free particle but has two possible orientations because the angle variable is not observable. A similar path is not possible for linear momentum. During each time interval Δ*t* there are two possible values of *l*=±*l*. One possible path in momentum space is shown above for the propagator *G(+l, t_f ;+l, t_i )*.

I note that this essentially one dimensional problem is embedded in three spatial dimensions with the axis of the one dimensional spin angular momentum taken to have arbitrary orientation. This propagator implies a two-state spinor that may be rotated to an arbitrary direction. A unitary rotation of the above propagator yields the propagator in three dimensions. This result is also the same as that found by calculating a time evolution operator. By using the momentum space path integral, a consistent theory results without the need to do the difficult task of constructing spin from unknowable angle variables [2], [8].

The above spin one-half fermion may be used to construct the propagator for a spin one boson. A particle composed of two identical spinning particles has only two momentum degrees of freedom, $l_1$ and $l_2$, in a one-dimensional system. The Hamiltonian is $H = \frac{1}{2I}l^2$, where now $l^2 = (l_1 + l_2)^2$, so that $\dot{l}_1 = -H_{\phi_1} = 0$ ($l_1 = \pm l_0$) and $\dot{l}_2 = -H_{\phi_2} = 0$ ($l_2 = \pm l_0$). Because the composite angular momentum *l* is the observable quantity, it is also constant during each time interval but now has three possible values, *+2l_0*, *-2l_0*, and *0*. The problem of a composite particle in three spatial dimensions or the problem of orbital angular momentum have Hamiltonians that couple angle and angular momentum variables. These problems could also be treated using this prescription. These more difficult problems are not treated here.

**Conclusion**

When extremized, the action *R* (*S*) results in Hamilton's equations for fixed beginning and ending momenta (position). Both of the actions *R* and *S* have a Hamilton-Jacobi equation (see Appendix A). *S* and *R* have some similarities and differences that were found by putting them into convex form and examining their second variation (see Appendix B and C). Using this parallel classical mechanics defined using the action *R*, the momentum path integral may be constructed from classical momentum paths and the momentum representation in quantum mechanics is shown to have its origin in classical mechanics. I show that this broader view of the path integral greatly improves this formulation of quantum mechanics by eliminating some of the previous weaknesses. In particular, the amplitude for a free particle is found without using the Schrödinger equation and the internal spin degree of freedom is simply and naturally derived. It also helps in understanding why Canonical transformations do not work in the phase space path integral.

I gratefully acknowledge the discussions with Arun Roy and the hospitality of Yves Garrabos at the ICMCB. I also gratefully acknowledge and the financial support from the Université de Bordeaux.

**Appendix A**

I now show that the classical mechanics defined from the functional *R* has a Hamilton-Jacobi equation. In the following I review results for the action *S* and present new results for the action *R*. The total differential of any arbitrary function *F(q, Q, t)* can be added to a Canonically transformed Lagrangian, i.e.,

$$\delta S = \delta\left\{\int_{t_i}^{t_f}(p\dot{q} - H(p,q))dt\right\} = \delta\left\{\int_{t_i}^{t_f}\left(P\dot{Q} - H^*(P,Q) + \frac{dF(q,Q,t)}{dt}\right)dt\right\} = 0$$

because

$$\delta\left[\int_{t_i}^{t_f}\left(\frac{dF(q,Q,t)}{dt}\right)dt\right] = \delta[F(t_f) - F(t_i)] = 0.$$

Equating the above integrands yields:

$$\left(p - \frac{\partial F}{\partial q}\right)\dot{q} - H(p,q) = \left(P + \frac{\partial F}{\partial Q}\right)\dot{Q} - H^*(P,Q,t) + \frac{\partial F(q,Q,t)}{\partial t}$$

or

$$p = \frac{\partial F(q,Q,t)}{\partial q}, -P = \frac{\partial F(q,Q,t)}{\partial Q}, H^*(P,Q,t) = H(p,q) + \frac{\partial F(q,Q,t)}{\partial t}.$$

If the coefficients of the generalized velocity are zero, then $\delta S = 0$, as can be seen in the above equation, and this implicitly defines a Canonical transformation. The function $F$ is said to generate a Canonical transformation.

The above argument only works for $F = F(q,Q,t)$. I can, however, Legendre transform from $(q,Q)$ to $(q,P)$ by defining a new generator $S = PQ - F$. After the usual manipulations I find that $S = S[q,P]$ and

$$p = \frac{\partial S(q,P,t)}{\partial q}, Q = \frac{\partial S(q,P,t)}{\partial P}, \frac{\partial S(q,P,t)}{\partial t} = \frac{\partial F(q,Q,t)}{\partial t}.$$

In particular, I can transform to a coordinate system where $P$ and $Q$ are fixed points or *constant* by choosing an $S$ that makes $H^* = 0$. This condition, when I substitute the expression above for $p$, becomes an equation for $S$ known as the Hamilton-Jacobi equation

$$H\left(\frac{\partial S(q,P,t)}{\partial q}, q\right) + \frac{\partial S(q,P,t)}{\partial t} = 0.$$

This equation includes the constant $P$ that can be considered a constant of integration. In fact, the generator $S$ can be shown to be the action along the actual path [9]. So the action $S$ generates a Canonical transformation from the phase space $(p, q)$ where the path may have various geometrical properties to the phase space $(P, Q)$ where the dynamics is a fixed point. This fixed point is directly related to the integration constants or initial conditions of the motion in the original system. Since the initial conditions determine, in principal, the motion in the original phase space, no information is lost in this transformation but the system has been greatly simplified.

The above Hamilton-Jacobi equation can be solved for $S[q,P,t]$. The action can also be calculated if $q(t)$ is known from the original definition of $S$ and the initial conditions, $q(t_i)$ and $p(t_i)$. I note that the beginning and ending conditions needed for a Canonical transformations $\delta p(t_i) = \delta q(t_i) = \delta p(t_f) = \delta q(t_f) = 0$ are the same ones that also allow both $S$ and $R$ to be simultaneously stationary. In fact, I can do a similar analysis for $R$ if I define the function $R$ by $F = pq + R$ implying that

$$dF = pdq + qdp + dR = \frac{\partial F(q,Q,t)}{\partial q}dq + \frac{\partial F(q,Q,t)}{\partial Q}dQ + \frac{\partial F(q,Q,t)}{\partial t}dt,$$

$$dR = -qdp - PdQ + \frac{\partial F(q,Q,t)}{\partial t}dt,$$

$$q = -\frac{\partial R(p,Q,t)}{\partial p}, P = -\frac{\partial R(p,Q,t)}{\partial Q}, \frac{\partial R(p,Q,t)}{\partial t} = \frac{\partial F(q,Q,t)}{\partial t},$$

and

$$H^*(P,Q,t) = H(p,q) + \frac{\partial R(p,Q,t)}{\partial t}.$$

Selecting an $R$ that makes $H^*=0$, I get another Hamilton-Jacobi equation for $R$

$$H\left(p, -\frac{\partial R(p,Q,t)}{\partial p}\right) + \frac{\partial R(p,Q,t)}{\partial t} = 0.$$

The $R$ function that solves this equation can easily be interpreted after taking the total time derivative, i.e.,

$$\frac{dR(p,Q,t)}{dt} = \frac{\partial R(p,Q,t)}{\partial p}\dot{p} + \frac{\partial R(p,Q,t)}{\partial t} = -q\dot{p} - H$$

or

$$R = -\int_{t_i}^{t_f}(q\dot{p} + H)dt = \int_{t_i}^{t_f}Kdt.$$

**Appendix B**

I know from statistical mechanics that the character of an extrema is extremely valuable knowledge in understanding a physical system. There are, however, some major difficulties associated with most variational principles especially when determining the type of extremum. To gain a broader view of the relation between $S$ and $R$, I now begin examining the types of bounds. One type of Hamiltonian that results in global bounds on the action is the saddle function Hamiltonian [10]. In this case, $H$ is a minimum in $p$ and a maximum in $q$. In the following I review results for the bounds on the action $S$ when $H$ is a saddle function and present new results for the action $R$.

I can test whether a function is convex or concave over a given domain while relaxing the continuity requirements by using the follow definition of convexity: A function $H(q) \in R$ is *convex* on *(a, b)* if $H(\lambda q_1 + (1-\lambda)q_2) \leq \lambda H(q_1) + (1-\lambda)H(q_2)$ for $0 < \lambda < 1$ and any $q_1$ and $q_2$ in *(a, b)*. $R$ is the set of real numbers. A function is strictly convex if the inequality is strict for distinct $q_1$ and $q_2$. A function is concave if $-H$ is convex.

The above definition expresses the geometry of the 'chord above arc' ($z = \lambda H(q_1) + (1-\lambda)H(q_2)$ is the chord in $\lambda$ between $q_1$ and $q_2$ while $z = H(\lambda q_1 + (1-\lambda)q_2)$ is the arc). If I choose any two points $\{q_1, q_2\}$ within the domain $\{a, b\}$, then any value of the function $H(q)$ between $q_1$ and $q_2$ (i.e., $H(\lambda q_1 + (1-\lambda)q_2)$) is less then the chord that connects $H(q_1)$ and $H(q_2)$ (i.e., $\lambda H(q_1) + (1-\lambda) H(q_2)$). It is easily proven that if $H(q)$ is differentiable the above definition is equivalent to $H(q_1) - H(q_2) - (q_1-q_2)H'(q_2) \geq 0$ for any $\{q_1, q_2\}$ in $\{a, b\}$, where $H'(q_2) = dH(q_2)/dq$ [10]. For a twice differentiable function the usual convexity test is recovered, i.e., $H$ is convex in *(a, b)* if $H_{qq} \geq 0$ for all $q$ in *(a, b)*.

The convex function definition can be extended to functions of two or more independent variables. The function $H(q,y)$ is convex with respect to $q$ if $H(\lambda q_1+(1-\lambda)q_2,y) \leq \lambda H(q_1,y) + (1-\lambda)H(q_2, y)$ for $0 < \lambda < 1$ and any $\{q_1, q_2\}$ in $\{a, b\}$. As in the above, if the partial derivative $H_q$ exists, then $H$ is convex with respect to $q$ if $H(q_1, y)-H(q_2, y) -(q_1-q_2) H_q(q_2,y) \geq 0$. This expanded definition also includes the saddle function case; if a function $H(q,y) \in R$ is convex in $q$ and concave in $y$, then it is a *saddle function*. A differentiable saddle function has both $H(q_1, y)-H(q_2, y) -(q_1-q_2)H_q(q_2,y) \geq 0$ and $H(q, y_1)-H(q, y_2) -(y_1-y_2)H_y(q, y_2) \leq 0$.

In reference [10] these definitions are applied to the functional $S$. $S$ is written in terms of the arbitrary momentum $\Pi$ and the position $\Theta$ to distinguish them from their critical functions *(p, q)* that extemize $S$, i.e., $S(\Pi, \Theta) = \int_{t_i}^{t_f} \{\Pi \dot{\Theta} - H(\Pi, \Theta, t)\} dt$ where the critical curve *(p,q)* satisfies Hamilton's equations $\dot{q} = \frac{\partial H}{\partial p} = H_p$ and $\dot{p} = -\frac{\partial H}{\partial q} = -H_q$. The beginning and ending conditions are $\Theta(t_i)-q(t_i)=\delta q(t_i)=0$ and $\Theta(t_f)-q(t_f)=\delta q(t_f)=0$. I can construct another functional from $S[\Pi,\Theta]$ by restricting $\Pi$ to be a solution to the equation $\dot{\Theta} = H_\Pi(\Pi, \Theta, t)$. Substituting the solution $\Pi=\Pi\,(\Theta)$ into $S$, I get the new functional $J[\Theta]=S[\Pi(\Theta),\Theta]$, and $J[q]=S[p,q]$ is the usual Euler-Lagrange form of the action. The difference is

$$J(\Theta) - J(q) = \int_{t_i}^{t_f} \{\Pi \dot{\Theta} - H(\Pi, \Theta, t) - p\dot{q} + H(p,q,t)\} dt.$$

I can put the integrand into the form of a convex or concave function by adding and subtracting $H(p, \Theta, t) - p\dot{\Theta}$ giving

$$J(\Theta) - J(q) = \int_{t_i}^{t_f} \{H(p, \Theta, t) - H(\Pi, \Theta, t) - (p - \Pi) H_\Pi\} dt$$

$$- \int_{t_i}^{t_f} \{H(p, \Theta, t) - H(p,q,t) - (\Theta - q) H_q\} dt - \left[p(q-\Theta)\right]_{t_i}^{t_f},$$

where I have used $\dot{\Theta} = H_\Pi$, partial integration, and $-\dot{p} = H_q$. If $\Theta$ is made to satisfy the initial and final conditions $\delta q(t_i)=0$ and $\delta q(t_f)=0$, i.e., $\Theta(t_i) = q(t_i)$ and $\Theta(t_f) = q(t_f)$, and $H(p,q,t)$ is a saddle function (i.e., it is convex in $p$ and concave in $q$), then the global minimum principle $J[q] \leq J[\Theta]$ holds.

These same conditions can be used to derive another global bound. In this case I define the functional $G[\Pi] = S[\Pi, \Theta(\Pi)]$ where $\Theta(\Pi)$ is the solution to $\dot{\Pi} = -H_\Theta(\Pi, \Theta, t)$. Following a similar argument (add and subtract $H(\Pi, q, t) - q\dot{\Pi}$, integrate by parts; use $\dot{q}=H_p$, $\dot{\Pi}=-H_\Theta$, $\delta q(t_i)=0$, and $\delta q(t_f)=0$) I get

$$G(p) - G(\Pi) = \int_{t_i}^{t_f} \{H(\Pi, q, t) - H(p, q, t) - (\Pi - p) H_p\} dt$$

$$- \int_{t_i}^{t_f} \{H(\Pi, q, t) - H(\Pi, \Theta, t) - (q - \Theta) H_\Theta\} dt.$$

If $H$ is a saddle function, then the global maximum principle $G[\Pi] \leq G[p]$ holds. Combining these results one can see that if $H$ is a saddle function and $\Theta$ satisfies the initial and final conditions $\Theta(t_i) = q(t_i)$ and $\Theta(t_f) = q(t_f)$ ( or $\delta q(t_i)=0$ and $\delta q(t_f)=0$), then
$$G(\Pi) \leq G(p) = S(p,q) = J(q) \leq J(\Theta),$$
where $\dot{\Theta} = H_\Pi$ is used to define $J(\Theta)$ and $\dot{\Pi} = -H_\Theta$ is used to define $G(\Pi)$.

I now do a similar analysis for the functional $R[\Pi,\Theta] = -\int_{t_i}^{t_f}(\Theta\dot{\Pi} + H)dt = \int_{t_i}^{t_f} K dt$ subject to the beginning condition $\delta p(t_i)=0$ and ending condition $\delta p(t_f)=0$. The functional $G'[\Pi] = \int_{t_i}^{t_f} K(\Pi, \Theta(\Pi))dt$ is constructed using the solution to the equation $\dot{\Theta} = H_\Pi$, i.e., $\Theta=\Theta(\Pi)$. As in the above case, $G'$ may be put in the form

$$G'(\Pi) - G'(p) = \int_{t_i}^{t_f}\{H(p,q,t) - H(p,\Theta,t) + (q-\Theta)\dot{p}\}dt$$

$$-\int_{t_i}^{t_f}\{H(\Pi,\Theta,t) - H(p,\Theta,t) - H(p,q,t) - (\Pi-p)\dot{\Theta}\}dt - +[p\Theta]_{t_i}^{t_f} - [\Pi\Theta]_{t_i}^{t_f}.$$

Using the above constraint $\dot{\Theta} = H_\Pi$, Hamilton's equation $\dot{p} = -H_q$, and the beginning and ending conditions $\delta p(t_i)=0$ and $\delta p(t_f)=0$, I get

$$G'(\Pi) - G'(p) = \int_{t_i}^{t_f}\{H(\Pi,q,t) - H(p,q,t) - (\Pi-p)H_p\}dt$$

$$-\int_{t_i}^{t_f}\{H(\Pi,q,t) - H(\Pi,\Theta,t) - (q-\Theta)H_\Theta\}dt,$$

i.e., $G'(p) \leq G'(\Pi)$ for a saddle function $H$. The functional $J'[\Theta] = \int_{t_i}^{t_f} K(\Pi(\Theta),\Theta)dt$ is defined using $\Pi(\Theta)$ that is a solution to $\dot{\Pi} = -H_\Theta$. In a similar manner, with the addition of an integration by parts, I obtain

$$J'(q) - J'(\Theta) = \int_{t_i}^{t_f}\{H(p,\Theta,t) - H(\Pi,\Theta,t) - (p-\Pi)H_\Pi\}dt$$

$$-\int_{t_i}^{t_f}\{H(p,\Theta,t) - H(p,q,t) - (\Theta-q)H_q\}dt,$$

i.e., $J'(\Theta) \leq J'(q)$.

Putting these together gives: $J'(\Theta) \leq J'(q) = R(p,q) = G'(p) \leq G'(\Pi)$. Note that these bounds are opposite in the sense that $R$ is bounded from *above* by a functional in $\Pi$ and *below* by a functional in $\Theta$ whereas $S$ is bounded *below* and *above* by similarly define functionals of $\Pi$ and $\Theta$. In conclusion, for the saddle function Hamiltonian the functional $S$ ($R$) is bounded from below (above) by an associated functional in $\Pi$ and is bounded from above (below) by an associated functional in $\Theta$.

**Appendix C**

The mathematical test for a maximum, minimum, etc., of a functional is analogous to the maximum and minimum tests for a function in elementary calculus. If the first variation (derivative) is zero at a path (point), then there is an extremum at that path (point) and the sign of the second variation (second derivative) reveals the type of extremum (minimum, maximum, inflection point, etc.) [11]. In analogy to a function, I can test for the type of extremum of $S$ by taking the second variation (i.e., the variation of the functional that results from the first variation) to obtain

$$\delta^2 S = \int_{t_i}^{t_f} \frac{1}{2!} \left\{ \frac{\partial^2 L}{\partial q^2} \delta q^2 + 2 \frac{\partial^2 L}{\partial q \partial \dot{q}} \delta q \delta \dot{q} + \frac{\partial^2 L}{\partial \dot{q}^2} \delta \dot{q}^2 \right\} dt.$$

As in elementary calculus, if $\delta S=0$ and $\delta^2 S>0$, independent of the choice of $\delta q$, then $S$ is a minimum for $q(t)$; or if $\delta S=0$ and $\delta^2 S<0$, independent of the choice of $\delta q$, then $S$ is a maximum for $q(t)$. If $\delta^2 S=0$ and all higher order variations are zero, then it is indeterminate for this method.

The bounds on $R$ and $S$, found in Appendix B, were determined by the properties of the Hamiltonian $H$. These bounds were found using restricted momentum or position variables so that the bounding functionals were dependent on only one conjugate variable. I now treat the $p$ and $q$ functions as independent, as in the derivation of Hamilton's equations, and study the type of extrema of $S$ and $R$. When $H$ is differentiable, I can express the second variation $\delta^2 S$ in terms of conjugate variables $(p,q)$ to get

$$\delta^2 S = \int_{t_i}^{t_f} \frac{1}{2!} \left\{ \frac{\partial^2 L}{\partial q^2} \delta q^2 + 2 \frac{\partial^2 L}{\partial q \partial p} \delta q \delta p + \frac{\partial^2 L}{\partial p^2} \delta p^2 \right\} dt.$$

The derivatives are evaluated at the critical $(p,q)$ and $L(p,q) = p\dot{q}(p,q) - H(p,q).$ In the Hamiltonian formulation the independent functions $p(t)$ and $q(t)$ are only connected by Hamilton's equations, and I may use $\dot{q} = H_p$ to get

$$\delta^2 S = \int_{t_i}^{t_f} \frac{1}{2!} \left\{ (\delta p, \delta q) \left\| \begin{array}{cc} H_{pp} + p H_{ppp} & p H_{ppq} \\ p H_{ppq} & p H_{pqq} - H_{qq} \end{array} \right\| \begin{pmatrix} \delta p \\ \delta q \end{pmatrix} \right\} dt = \frac{1}{2!} \int_{t_i}^{t_f} \left\{ V^T M V \right\} dt$$

where

$$M = \left\| \begin{array}{cc} H_{pp} + p H_{ppp} & p H_{ppq} \\ p H_{ppq} & p H_{pqq} - H_{qq} \end{array} \right\|, \quad V = \begin{pmatrix} \delta p \\ \delta q \end{pmatrix}. \qquad (1)$$

If I consider the problem $MV=\lambda V$, the symmetric matrix $M$ in equation (1) can be diagonalized yielding two real eigenvalues $\lambda_1$ and $\lambda_2$. The corresponding two eigenvectors are $V_1$ and $V_2$. The eigenvectors can be used to construct an orthogonal matrix $P$ ($P P^T = I$) that diagonalizes $M$, i.e., $P^T V = \begin{pmatrix} a \\ b \end{pmatrix}$ and $V^T P = (a, b)$. I find that

$$\delta^2 S = \frac{1}{2!} \int_{t_i}^{t_f} \left\{ V^T M V \right\} dt = \frac{1}{2!} \int_{t_i}^{t_f} \left\{ V^T P P^{-1} M P P^{-1} V \right\} dt = \int_{t_i}^{t_f} \left\{ (a,b) P^{-1} M P \begin{pmatrix} a \\ b \end{pmatrix} \right\} dt$$

$$= \int_{t_i}^{t_f} \left\{ (a,b) \begin{pmatrix} \lambda_1 & 0 \\ 0 & \lambda_2 \end{pmatrix} \begin{pmatrix} a \\ b \end{pmatrix} \right\} dt = \int_{t_i}^{t_f} \left\{ \lambda_1 (a)^2 \right\} dt + \int_{t_i}^{t_f} \left\{ \lambda_2 (b)^2 \right\} dt.$$

*a* and *b* are arbitrary because they consist of linear combinations of the independent variations δ*p* and δ*q*. The second variation will only have a constant sign if both eigenvalues have constant sign over the entire critical path as can be proven in the following. One of the two constants (e.g., *b*) can be made zero over the entire critical path while the other could be made to be zero over only a portion of the path. Any change of sign of $\lambda_1$ could make $\delta^2 S$ either sign by special choices of *a* that are zero while $\lambda_1$ is one sign and not zero when $\lambda_1$ is the other sign. A similar argument also works for $\lambda_2$. It just so happens that the conditions $\lambda_1>0$ and $\lambda_2>0$ for *S* to be a minimum (or $\lambda_1<0$ and $\lambda_2<0$ for *S* to be a maximum) are the same for when the matrix *M* is positive (negative) definite during $\Delta t = t_f - t_i$, i.e.,

$$H_{pp} + pH_{ppp} > (<) 0, \quad (H_{pp} + pH_{ppp})(pH_{pqq} - H_{qq}) - (pH_{ppq})^2 > 0, \qquad (4)$$

during $\Delta t$. If *M* is positive (negative) definite for all initial and final conditions, then *S* is a global minimum (maximum). This shows that the type of global extrema, if one exists, only depends on the form of *H*. This can be illustrated with the simple and practical Hamiltonian of the form $p^2/2m + V(q)$ making $H_{pp} > 0$ and $H_{ppp} = H_{ppq} = H_{pqq} = 0$. *S* is a global minimum if

$$H_{pp} > 0, \quad \begin{Vmatrix} H_{pp} & 0 \\ 0 & -H_{qq} \end{Vmatrix} = -H_{pp} H_{qq} > 0$$

for all $q(t_i)$ and $q(t_f)$. The second condition is only possible when $H_{qq}$ is negative (concave downward). This is an example of the saddle function case treated above, where I found upper and lower bounds on *S*. A concave downward potential is unstable about the maximum of the potential. The motion is unbounded where a particle accelerates away from this maximum, i.e., the motion approaches the "fixed point" of $p \to \pm\infty$ and $q \to \pm\infty$. In this case, *S* is a global minimum for a globally unstable and repulsive potential. The bounded motion for a concave upward potential (stable and attractive potential) is neither a global minimum nor a global maximum of the action. In this and more complex cases, the potential may have domains where $H_{qq}$ is positive (stable attractive region) and other domains where $H_{qq}$ is negative (an unstable or repulsive domain) over given time intervals. When *H* is locally a saddle function and $\Delta t$ is sufficiently small, the second condition is satisfied. It is also possible that it is equal to zero so that this method is indeterminate. These points are analogous to the critical points in thermodynamics [12].

In the more general case where *p* is at most second order, corresponding to the quadratic kinetic energy term, the action has a minima when $(pH_{pqq} - H_{qq})(H_{pp}) - (pH_{pqq})^2 > 0$. The type of extrema clearly depends on the details of a particular Hamiltonian. I can, e.g., write $A(q)p^2 + B(q)p + V(q)$ giving a global minima when $p^2(AA_{qq} - 4A_{qq}^2) \rangle AV_{qq}$. This condition does not depend on the *B(q)*.

It is easy to show that a Canonical transformation does not necessarily preserve the extrema character of the action. The action corresponding to the Hamiltonian $H = p^2 - q^2$ is a global minimum by equation (4). Using the generator $F = q^2 Q$ gives a totally indeterminate action, whereas $F = Q/q$, for a domain not including the origin, gives an action that is neither a maximum nor a minimum.

Following the discussion above, I can also show the condition for the type of extrema of *R*. The second variation is:

$$\delta^2 R = \int_{t_i}^{t_f} \frac{1}{2!} \left\{ \frac{\partial^2 K}{\partial q^2} \delta q^2 + 2 \frac{\partial^2 K}{\partial q \partial p} \delta q \delta p + \frac{\partial^2 K}{\partial p^2} \delta p^2 \right\} dt.$$

Using $\dot{p} = -H_q$, as above, I get

$$\delta^2 R = \int_{t_i}^{t_f} \frac{1}{2!} \left\{ (\delta p, \delta q) \begin{Vmatrix} qH_{ppq} - H_{pp} & qH_{qqp} \\ qH_{qqp} & qH_{qqq} + H_{qq} \end{Vmatrix} \begin{pmatrix} \delta p \\ \delta q \end{pmatrix} \right\} dt.$$

By the same argument as before, $R$ is minimum (maximum) if the matrix is positive (negative) definite during $\Delta t$. The Hamiltonian $H = p^2/2m + V(q)$ has $H_{pp} > 0$ and $H_{ppq} = H_{qqp} = 0$ so that the off diagonal elements are zero. $R$ is a maximum if

$$H_{pp} > 0, \quad \begin{Vmatrix} -H_{pp} & 0 \\ 0 & qH_{qqq} + H_{qq} \end{Vmatrix} = -H_{pp}(qH_{qqq} + H_{qq}) > 0.$$

The second condition is only possible when $(qH_{qqq}+H_{qq})$ is negative as occurs in a concave downward potential or a saddle function Hamiltonian. It is revealing that the inertial part of the Hamiltonian $H=p^2/2m + V(q)$ allows a minimum for $S$ and a maximum for $R$. This is similar to the results of Appendix B, where the bounds were found by eliminating one of the Canonical variables using one of Hamilton's equation.

It may be possible with more complicated and realistic Hamiltonians to categorize the phase space into domains identified by the extrema properties. These complications are even more evident in higher dimension where the diagonal elements of the *2nx2n* Hessian matrix $M$ have the form $\sum_{i=1}^{n} (p_i H_{p_i} - H)_{p_i p_i}$, and the off-diagonal elements are also generalized. Extending the above argument, the critical path is a minimum (maximum) if the eigenvalues of $M$ are all positive (negative) during $\Delta t$.